\documentclass[prx,longbibliography,showpacs,twocolumn,superscriptaddress,amsmath,amssymb,verbatim]{revtex4-2}
\usepackage{amsmath,amssymb,amsfonts,stmaryrd,wasysym,graphicx,multirow,color,textcomp,subfigure}
\usepackage{url}
\usepackage[colorlinks=true,citecolor=blue,urlcolor=blue]{hyperref}
\usepackage{romannum}
\usepackage{xcolor}
\usepackage{physics}
\usepackage{ulem}
\usepackage{soul}

\normalfont
\def\be{\begin{equation}}
	\def\ee{\end{equation}}
\def\bea{\begin{eqnarray}}
	\def\eea{\end{eqnarray}}

\begin{document}

 \title{Boundary-Driven Complex Brillouin Zone \\in Non-Hermitian Electric Circuits}
	
\author{Yung Kim}
\thanks{These authors contributed equally to this work.}
\affiliation{%
 Department of Physics, Korea Advanced Institute of Science and Technology, Daejeon 34141, Republic of Korea
}%
\affiliation{Department of Mechanical Engineering, Korea Advanced Institute of Science and Technology, Daejeon 34141, Republic of Korea}

\author{Sonu Verma}
\thanks{These authors contributed equally to this work.}
\affiliation{ Center for Theoretical Physics of Complex Systems, Institute for Basic Science, Daejeon 34126, Republic of Korea}

\author{Minwook Kyung}
\thanks{These authors contributed equally to this work.}
\affiliation{%
 Department of Physics, Korea Advanced Institute of Science and Technology, Daejeon 34141, Republic of Korea
}%

\author{Kyungmin Lee}
\affiliation{%
 Department of Physics, Korea Advanced Institute of Science and Technology, Daejeon 34141, Republic of Korea
}%

\author{Wenwen Liu}
\affiliation{Department of Electrical and Electronic Engineering, University of Hong Kong, Hong Kong, China}

\author{Shuang Zhang}
\affiliation{Department of Electrical and Electronic Engineering, University of Hong Kong, Hong Kong, China}

\author{Bumki Min}
  \email{bmin@kaist.ac.kr}
\affiliation{%
 Department of Physics, Korea Advanced Institute of Science and Technology, Daejeon 34141, Republic of Korea
}%
\affiliation{Department of Mechanical Engineering, Korea Advanced Institute of Science and Technology, Daejeon 34141, Republic of Korea}

\author{Moon Jip Park}
  \email{moonjippark@hanyang.ac.kr}
\affiliation{%
 Department of Physics, Hanyang University, Seoul 04763, Republic of Korea
}%

	\begin{abstract}

Complex-valued physical quantities, often non-conserved, represent key phenomena in non-Hermitian systems such as dissipation and localization. Recent advancements in non-Hermitian physics have revealed boundary-condition-sensitive band structures, characterized by a continuous manifold of complex-valued momentum known as the generalized Brillouin zone (GBZ). However, the ability to actively manipulate the GBZ and its associated topological properties has remained largely unexplored. Here, we demonstrate a controllable manipulation of the GBZ by adjusting the boundary Hamiltonian and leveraging the boundary sensitivity in a circuit lattice. Our observations reveal that the GBZ forms multiple separated manifolds containing both decaying and growing wave functions, in contrast to the previously observed non-Hermitian skin effect under open boundary condition (OBC). By continuously deforming the GBZ, we observe the topological phase transitions of innate topological structure of GBZ that are enriched by complex properties of non-Hermitian physical variables. Notably, such topological phase transition is governed by boundary conditions rather than bulk properties, underscoring the extreme boundary sensitivity unique to non-Hermitian systems.
	\end{abstract}
 
	\maketitle

Controlling the spatiotemporal propagation of waves is a fundamental interest in many areas of physics \cite{joannopoulos1997photonic,russell2003photonic,RevModPhys.91.015006, yang2024non,RevModPhys.76.323}. Recently, non-Hermitian systems have shown distinguished properties for this purpose, exhibiting highly sensitive band structures that respond to boundary conditions \cite{PhysRevLett.121.086803,PhysRevB.97.121401,zhang2022universal,PhysRevLett.125.180403, ding2022non, PhysRevB.105.245143}. This pronounced boundary sensitivity is a universal feature of wave phenomena, spanning classical to quantum systems, and is observed in platforms such as photonic \cite{weidemann2020topological, xiao2020non, PhysRevLett.132.063804, PhysRevLett.133.070801}, acoustic \cite{Zhang2021acoustic, s41467-021-26619-8, zhou2023observation, gu2022transient}, circuit \cite{helbig2020generalized, zhou2021, ochkan2024non, yuan2023non, zhu2023higher,wu2022non,hadad2018self}, and mechanical systems \cite{wang2022non, Ghatak2020, PhysRevResearch.2.013058}, as well as ultracold atomic gases \cite{PhysRevLett.129.070401, wang2023observation}.

\begin{figure*}[p]
\centering
	\includegraphics[width=0.9\textwidth]{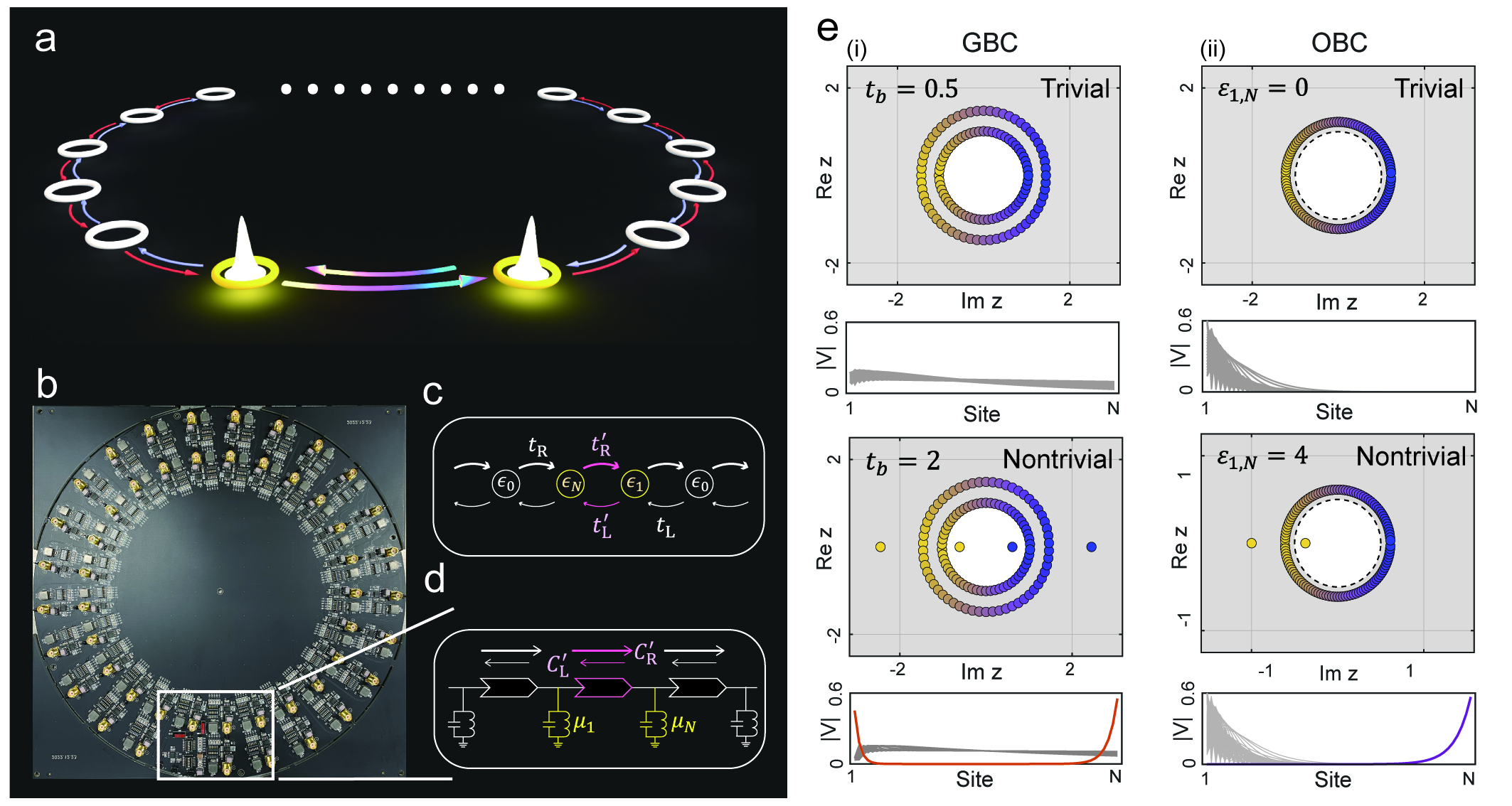}
	\caption{\textbf{Hatano-Nelson circuit with generalized boundary.} \textbf{a} Schematic illustration of the one-dimensional non-Hermitian HN model with GBC.\textbf{b} Photograph of the implemented printed circuit board (PCB).\textbf{c} Tight-binding model of our system. The parameters $\varepsilon_0, t_{\rm R}, t_{\rm L}$ represent onsite potential, hopping amplitudes for right and left direction of bulk sites. $\varepsilon_{\rm 1}, \varepsilon_{\rm N}, t'_{\rm R}, t'_{\rm L}$ denote the boundary onsite potentials and hopping amplitudes. \textbf{d} Corresponding circuit model. Non-reciprocal hoppings are realized by INICs. At the boundary, we implemented tunable switches to change the onsite potentials $C'_{\rm g}, L'_{\rm g}$ and hopping amplitudes $C'_{\rm R}, C'_{\rm L}$. \textbf{e} Topological phase transition of GBZ under GBC (i) and OBC (ii). The topological phase transition occurred depending on the changes of onsite potential and coupling amplitude, leading to the appearance of topological boundary modes in complex and real space simultaneously. Note that onsite potentials are fixed at $\varepsilon_{1,N}=0$ under GBC.}
	\label{figs:fig1}
\end{figure*}

Non-Hermiticity extends the real-valued momentum of the Brillouin zone (BZ) into a complex manifold, known as the generalized Brillouin zone (GBZ) under OBC \cite{PhysRevLett.121.086803, PhysRevLett.123.066404,PhysRevLett.125.126402,PhysRevLett.125.226402}. Given that boundary conditions in non-Hermitian systems are not limited to periodic or open forms \cite{PhysRevLett.127.116801, PhysRevResearch.5.033058, PhysRevA.104.022215}, the introduction of a continuously deformable boundary Hamiltonian allows a feasible approach to engineering non-Hermitian band structures without a holistic change to the system \cite{verma2024topological, guo2024scale, s42005-021-00547}. Here, we propose a circuit lattice with tunable boundary components as a promising framework for investigating boundary-sensitive phenomena in non-Hermitian systems (Fig.\ref{figs:fig1}(a)). We demonstrate experimentally that deforming the boundary Hamiltonian not only reshapes the morphology of the GBZ manifold but also induces changes in its topological structure, revealing a deep interplay between boundary conditions and system topology.

Consider the scattering matrix approach that relates the incoming and outgoing waves as they interact with the scattering source:
\\
\begin{equation}
\bold{\Psi}_{\textrm{out},k}=
\hat{S}_{kk'}\bold{\Psi}_{\textrm{in},k'},
\label{eq:scattering}
\end{equation}
\\
where $\bold{\Psi}_{\textrm{out}(\textrm{in}),k}$ are the amplitude of incoming (outgoing) wave functions with momentum $k$. Eq. (1) encodes the symmetry of the physical systems. For instance, in time-reversal symmetric medium, the scattering matrix is symmetric, $\hat{S}=\hat{S}^T$. With non-Hermiticity, scattering matrix is generally non-unitary, $\hat{S}^\dagger \hat{S}\neq I$, indicating the eigenvalues can extend outside the unit circle in the complex plane. 

Eq.\eqref{eq:scattering} also describes the lattice model with the boundary terms. For instance, in Hatano-Nelson (HN) model \cite{hatano1996localization} with a boundary, the wave packets with distinct momenta are coupled by the scattering matrix as,
\\
\begin{align}
	\label{eqs:H_b}
 \left[ 
\begin{array}{c}
	|\Psi_\textrm{R} (z_1)\rangle\\
	|\Psi_\textrm{L} (z_2)\rangle 
 \end{array}
 \right]
 =
\begin{bmatrix}
	T_{11} & R_{12} \\
	R_{21} & T_{22}
	\end{bmatrix} 
 \left[ 
\begin{array}{c}
	|\Psi_\textrm{L} (z_1)\rangle\\
	|\Psi_\textrm{R} (z_2)\rangle 
 \end{array}
 \right],
	\end{align}
    \\
where $|\Psi_\textrm{L(R)} (z)\rangle$ is the wave function of the left (right) side respect to the boundary. The complex momentum is written as, $z=e^{ik}$. Here, we denote $z_1=e^{ik}$ and $z_2=e^{-ik}$, which define the forward and backward wave propagation, respectively. Therefore, $T_{ij}$ and $R_{ij}$ with $i,j\in\lbrace 1,2\rbrace$ are the transmission and reflection amplitudes, respectively. The relation between scattering matrix theory and non-Bloch band theory is illustrated in supplementary material (SM).
 
Our circuit system is designed to emulate the non-Hermitian lattice model with tunable boundary Hamiltonian. The current-voltage response of the circuit system is given as, $\textbf{I}=\hat{J}\textbf{V}$, where $\hat{J}$ is the admittance matrix (or circuit Laplacian)  \cite{lee2018topolectrical}. Here, $\textbf{I}$ and $\textbf{V}$ represent the vector components of the input current and the output node voltage response, respectively. $\hat{J}$ plays a role of the tight-binding Hamiltonian, where this correspondence enables direct access to physical information about the eigenstate and eigenvalue spectrum \cite{PhysRevB.99.161114}. This allows us to extract the eigenvalues of $\hat{J} $ through measurements of the admittance spectrum and voltage distributions across circuit nodes.

The circuit Laplacian is composed of the bulk and the controllable boundary term as, $\hat{J} = \hat{J}_{\textrm{bulk}} + \hat{J}_{\textrm{bdy}}$ (Fig. 1a-d): 
\\
\begin{equation}
\begin{aligned}
    J_{\textrm{bulk}}&= \sum_{n=2}^{N-1} 2C_1|n\rangle\langle n| +\sum_{n=1}^{N}\left(C_{\rm g} - \frac{1}{\omega^2L_{\rm g}} \right) |n\rangle\langle n| \\
    & +\sum_{n=1}^{N-1} \left(C_{\rm L}|n\rangle\langle n+1| + C_{\rm R}|n+1\rangle\langle n|\right)
\end{aligned}    
\label{EQs:1}
\end{equation}
\\
and
\\
\begin{equation}
    \begin{aligned}
       & J_{\textrm{bdy}}= \mu_1|1\rangle\langle 1| + \mu_N|N\rangle\langle N| 
         - C_{\rm R}'|N\rangle\langle 1| - C_{\rm L}'|1\rangle\langle N|,
    \end{aligned}
    \label{EQs:2}
\end{equation}
\\
where $C_{\rm R}$ and $C_{\rm L}$ represent the coupling capacitances corresponding to the right and left hopping amplitudes between bulk nodes, $C_{\rm g}$ and $L_{\rm g}$ represent the capacitance and inductance for the ground resonance of each node. At the boundary nodes, the onsite potentials are defined as $\mu_1 = C_{\rm L} + t_{\rm b} C_{\rm R} + \varepsilon_1$ and $\mu_N = C_{\rm R} + t_{\rm b} C_{\rm L} + \varepsilon_N$. The boundary coupling capacitances are $C'_{\rm R} = t_{\rm b} C_{\rm R}$ and $C'_{\rm L} = t_{\rm b} C_{\rm L}$, which represent the non-reciprocal hopping amplitudes between the boundary nodes. Here, $\varepsilon_{\rm 1,N}$ and $t_{\rm b}$ are boundary parameters that can be modified using adjustable switches. Each node in the circuit is connected through capacitors in parallel with an operational amplifier, also known as negative impedance converter with current inversion (INIC \cite{PhysRevLett.122.247702}), enabling non-reciprocal hopping amplitudes. The detailed derivation of the circuit Laplacian is provided in the SM.  

The non-reciprocal hopping causes the GBZ to deviate from the unit circle, forming contours that satisfy $z_1 z_2=t_\textrm{R}/t_\textrm{L}$ (OBC in Fig. 1e), where $t_\textrm{L}$ and $t_\textrm{R}$ are the hopping amplitude for the left and right direction. Under OBC, this deviation of the GBZ from the unit circle signifies macroscopic boundary localization, commonly referred to as the non-Hermitian skin effect. However, when the boundary hopping is not fully terminated—referred to here as generalized boundary conditions (GBC)—we observe the formations of the separated manifold in the GBZs, a feature absent in both periodic boundary conditions (PBC) and OBC.  The resulting eigenstates can be expressed as a linear combination of wave packets with distinct complex momenta and radii, as follows:
\\
 \bea
|\Psi \rangle=c_1 |\Psi(z_1)\rangle +c_2 |\Psi(z_2)\rangle,
\label{EQs:5}
 \eea
 \\
where $|z_1|\neq|z_2|$. The non-Hermitian nature of the system, specifically non-reciprocal hopping, causes a unidirectional propagation. Consequently, in the presence of boundary scatterers, the system exhibits wave packets with transmission probability ($|T_{ii}|^2$ in Eq. 2) differing from unity. These states define the separation of manifolds in the GBZs as shown in the left panels of Fig. 1e. 

Fig. 1e illustrates the eigenstates in the case of GBCs compared to OBC. Due to the separated GBZ, the eigenstates exhibit accumulated envelopes on both the left and right sides of the boundary. Microscopically, finite hopping between boundary sites enables the coexistence of left- and right-evanescent waves. This bidirectional localization of the eigenstates at the boundaries arises from the interplay between the non-Hermitian skin effect and boundary conditions. While non-reciprocal hopping alone drives wave accumulation at one edge (the non-Hermitian skin effect), the introduction of finite boundary potentials or modified boundary couplings creates additional localized states that overlap with the skin modes, giving rise to bidirectional boundary localization.
In the absence of non-reciprocal hopping, the boundary term can trap at most two bound states at the boundary, similar to typical bound states in Hermitian systems \cite{RevModPhys.83.1057}. On the other hand, the presence of non-reciprocal hopping allows the non-orthogonality of the eigenstates, which allows for the macroscopic accumulation of states at the boundary and localization across more than two sites.
The separated GBZ possesses inherent topological structures \cite{verma2024topological}. The coupled momenta in Eq. (\ref{EQs:5}) can be recast into an effective spinor form, $ \psi_n(z_1,z_2) = c_1z_1^n + c_2 z_2^n$, which includes a well-defined relative phase, $ \phi = \textrm{arg}(c_2 / c_1)$, between the two wave components. The meromorphic function $c_2/c_1$ is determined by the two boundary equations $c_2/c_1=1/h_{\rm B}^+(z_1, z_2)=h_{\rm B}^-(z_1, z_2)$. It is related to the reflection and transmission coefficients in Eq. 2 as $(c_1,c_2)$ determines the amplitudes of incoming and outgoing wave packets [see Eq. (5) and SM]. Therefore, a well-defined phase of $c_2/c_1$ allows for the definition of a winding number, which serves as the topological invariant of the GBZ:
\\
\bea
W=\frac{W_{+}-W_{-}}{2},~W_{\pm}=\frac{1}{2\pi}[\textrm{arg}(h_{\rm B}^{\pm}(z_1, z_2))]_{\mathcal{L}_1\times\mathcal{L}_2},
\eea
\\
here $\mathcal{L}_1\times\mathcal{L}_2$ indicates the continuous contour of the GBZs satisfying $z_1z_2 = t_{\rm R}/t_{\rm L}.$ $\arg[h_{\rm B}^{\pm}(z_1, z_2)]_{\mathcal{L}_1\times\mathcal{L}_2}$ is the change of phase of $h_{\rm B}^{\pm}(z_1, z_2)$ as $(z_1, z_2)$ goes along the GBZ.
The nontrivial winding number manifests the emergence of topological boundary modes, as illustrated in Fig. 1e. Eq.~(5) can be interpreted as an effective spinor representation of the wave functions analogous to the SSH model, where the incoming and outgoing wave functions (or wave packets) play the role of the effective sublattice degrees of freedom.

To experimentally reconstruct the GBZ, we employ the wave function projection method by projecting the eigenstates of the system onto the complex plane using a complex Fourier transform (See SM). By tuning the boundary parameters $t_{\rm b}, \varepsilon_{1,N}$ using 8 way dip switches (See SM), we observe the corresponding topological phase transitions, marked by the appearance or disappearance of the topological boundary modes. In the nontrivial phase, the topological boundary mode appears as distinct, separated from the complex manifolds of the bulk momenta. In contrast, in the trivial phase, both GBC and OBC exhibit only bulk modes with no topological boundary modes. As the GBZ undergoes the topological phase transition, the complex momenta of the boundary state are absorbed into the bulk spectrum. This absorption coincides with an exceptional point (EP), directly indicating the coalescence of the wave functions. Notably, this coalescence differs from the energy band touchings, as energy degeneracy does not necessarily imply the coalescence of eigenstates. 

 Figure 2a-d presents a comprehensive analysis of how boundary parameters, $t_{\rm b}$ and $\varepsilon$, influence the topological properties and the corresponding physical manifestations in our system. Here, $\varepsilon$ is defined as $\varepsilon \equiv \varepsilon_1 = 0.5\varepsilon_N$ to clearly reveal phase transitions. It is important to note that the parameters used in these calculations are retrieved from the measured circuit Laplacian, ensuring consistency between theory and experiments. In Fig. 2a, the non-Bloch winding number $W$ serves as a topological invariant, characterizing the system's topological phases, while the number of bound states in Fig. 2b directly corresponds to the emergence of topological boundary modes, a key observable in our setup. Additionally, the phase rigidity $R$ in Fig. 2c quantifies the biorthogonality of the eigenstates and the inverse participation ratio (IPR) in Fig. 2d indicates the localization strength of the eigenstates. Notably, figures 2a-d reveal distinct topological phase regions in the boundary parameter space, clearly demonstrating the critical role of boundary matrix in determining topological phases. 
 
 The occurrence of topological phase transitions at EPs is one of the key features of our system. At these critical points, the GBZs make contact, resulting in the coalescence of eigenstates. Unlike conventional energy band touchings in Hermitian systems, this merging of eigenstates at EPs leads to a distinct class of non-equilibrium phase transitions, known as exceptional transitions \cite{fruchart2021non}. This phenomenon can be quantitatively identified through the phase rigidity, which measures the biorthogonality of eigenstates:
 \\
\bea
R_i = \frac{\langle \psi^{\rm R}_i | \psi^{\rm L}_i \rangle}{\langle \psi^{\rm R}_i | \psi^{\rm R}_i \rangle} \quad \text{for}  \quad i=1,2,\ldots,N.
\eea
\\
Here, $|\psi_{i}^{\rm R (L)}\rangle $ denotes the right (left) eigenstate. Figure 2c maps the average phase rigidity $(R_1+R_2+R_{N-1}+R_N)/4$ for topological boundary states, showing that it approaches zero at topological phase transition points, providing a clear signature of exceptional transitions. To quantify the boundary localization of topological boundary states, we use the IPR \cite{PhysRevB.97.121401}:
\\
\begin{equation}
    \text{IPR}_i = \frac{\sum_\mathbf{r} | \psi_i(\mathbf{r}) |^4 \,}{\left( \sum_\mathbf{r} | \psi_i(\mathbf{r}) |^2 \,\right)^2} \quad \text{for} \quad i=1,2,\ldots,N.
\end{equation}
\\
and Fig. 2d illustrates the IPR of topological states with the color gradient indicating the degree of eigenstate localization. This analysis reveals that topological states of the GBZ exhibit boundary localization, similar to conventional topological states. Furthermore, Fig. 2h shows an inverse relationship between the localization of topological boundary states in real space and their spread in the complex momentum space. As these states become more localized in real space, their distribution in the complex momentum space widens correspondingly (See SM).

Figure 2g shows a comparison of theoretically calculated and experimentally measured GBZs. Despite the presence of disorder in the circuit lattice such as parasitic values in lumped elements and imperfection of fabrication, the peaks in the experimental GBZ still closely align with the theoretical predictions, confirming the robustness of the GBZ topology even in the presence of moderate disorder. Figure 2h presents the experimentally measured GBZ under both OBC (upper panels) and GBC (lower panels). The distribution peaks in the GBZ align well with our theoretical predictions. 

In order to present quantitative evidence for the occurrence of topological phase transitions, we demonstrate the direct measurements of the phase rigidity and IPR of the topological boundary states. Figure 2e,f illustrate these experimental results under GBC and OBC. For GBC, we fixed the $\varepsilon$ and varied $t_{\rm b}$ (yellow dashed line in Fig. 2b). Conversely, for OBC, we fixed $t_{\rm b} = 0$ and adjusted the $\varepsilon$ (white line in Fig. 2b). In both boundary conditions, we observed two significant dips in the phase rigidity plot corresponding to the EPs shown in Fig. 2c. Although phase rigidity did not reach zero due to system size limitations, it approached values near zero, indicating the proximity to EPs. Furthermore, under the same physical conditions, the IPR of the topological boundary states approached 1, signifying strong boundary localization. These observations collectively support the occurrence of topological phase transitions in our system. The specific boundary parameters used here are provided in Section C of the SM.

In addition to the Hatano-Nelson model, we implement the non-reciprocal next nearest-neighbor(NNN) hopping by implementing additional INICs with capacitors $C_{\rm R}$ and $C_{\rm L}$ at each node. As shown in Fig. 3a, these extra couplings replicate the same magnitude as NN couplings to realize the NNN hoppings. Details of circuit Laplacian are provided in SM. In the presence of the NNN hoppings, the bulk equation for the momenta is given as quartic equation of momentum, $t_{2\rm L} z^4 + t_{1\rm L} z^3-E z^2+t_{1\rm R} z+t_{2 \rm R} = 0$, here $t_{1\rm R} (t_{1\rm L})$ and $t_{2\rm R} (t_{1\rm L}) $ denotes the nearest neighbor right (left) and next nearest neighbor right (left) hopping amplitudes, respectively. In general, the bulk equation has four generalized momenta solutions $z_i,~i=1,2,3,4$, and the resulting eigenstates are given by: 
\\
 \bea
|\Psi \rangle=c_1 |\Psi(z_1)\rangle +c_2 |\Psi(z_2)\rangle+c_3 |\Psi(z_3)\rangle + c_4 |\Psi(z_4)\rangle,
 \eea
 \\
where $z_i, c_i,~i=1,2,3,4$ are determined self-consistently by solving bulk and boundary equations. The introduction of NNN hoppings not only reveals a more complicated topological structure of the GBZ, but also leads to multiple topological states under GBC and OBC as shown in Fig. 3b \cite{PhysRevLett.125.226402}. The experimentally measured GBZs for trivial and nontrivial phases under GBC and OBC are shown in Fig. 3c-f. Our theoretical model reveals a complicated structure, but in practical experiments, particularly when employing  Fourier transform, only the most prominent peaks are present. This leads to the manifestation of distinctive, simplified patterns in the complex plane, a phenomenon clearly illustrated in Fig. 3c-f. Note that the topological boundary states, as observed in both real and complex planes, demonstrate behaviors parallel to those found in nearest-neighbor (NN) models. This resemblance implies that the localization properties in NN and long-range hopping models share fundamental characteristics in the context of the intrinsic topology of GBZ.

To summarize, we have demonstrated the manipulation of the GBZ using non-Hermitian scatterers and visualize its topological structure within a non-Hermitian circuit system. While several studies have examined spectral changes in the OBC of the non-Hermitian skin effect, here we find that, by precisely controlling the boundary Hamiltonian, the two momenta form separate manifolds, extending the scope of engineering the GBZ manifold in a controllable fashion, retaining the relative phase and amplitude information between the two manifolds. To further generalize our understanding of GBZ topology, we introduce next-nearest-neighbor hopping into the system. This extension leads to the observation of multiple topological boundary modes within the complex momentum space, enriching our understanding of non-Hermitian topological systems. Our findings provide the much-needed experimental evidence that the observed phenomena stem from the intrinsic topological properties of the GBZ itself, bridging the gap between theory and practice, and opening new avenues for investigating and manipulating topological properties in non-Hermitian materials. 

\bibliography{reference}

\begin{figure*}[p]
  \centering
  \includegraphics[width=0.9\textwidth]{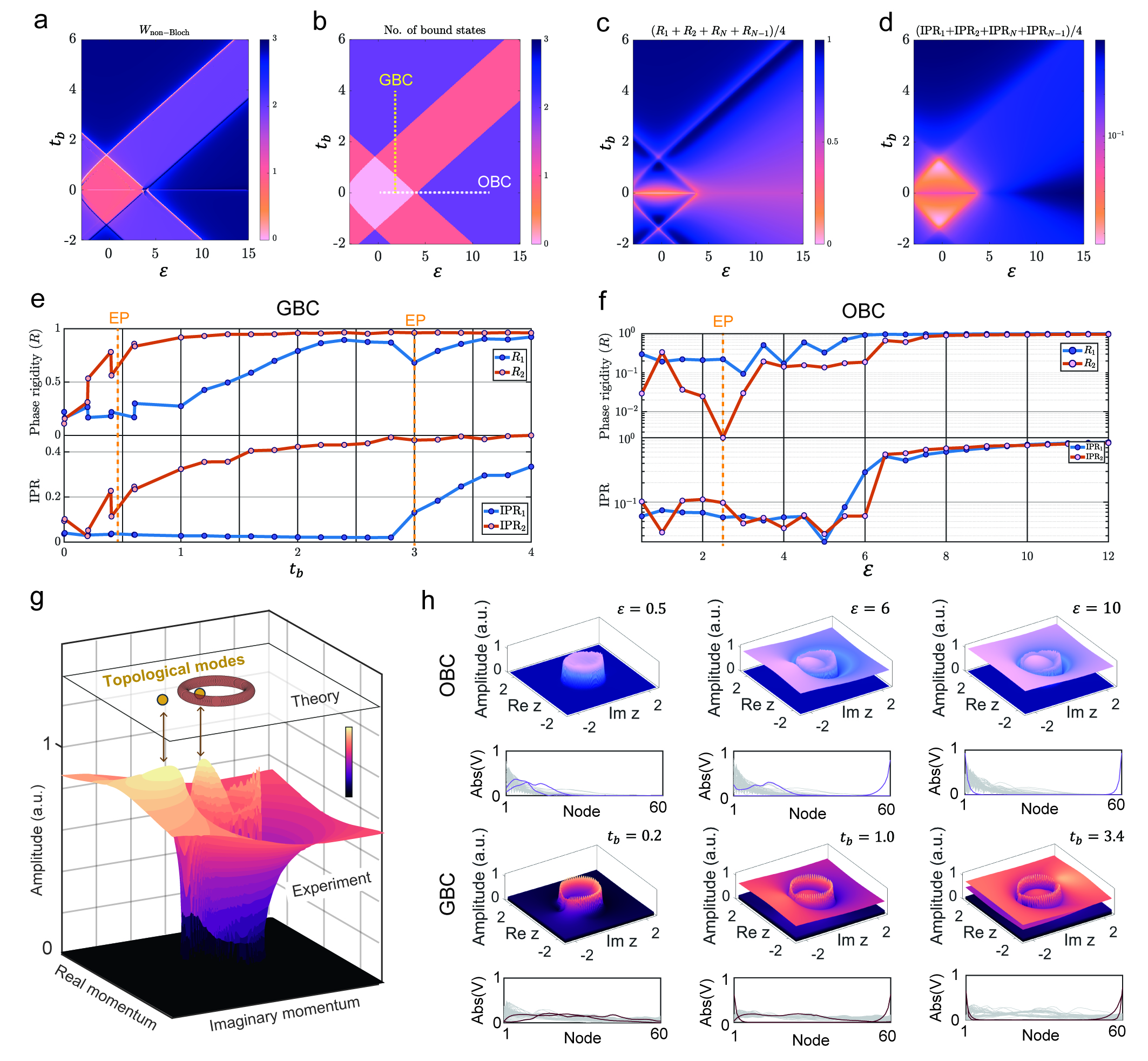}
  \caption{\textbf{Intrinsic topology of GBZ.} \textbf{a-d} Phase diagram of non-Blcoh winding number, number of boundary states, phase rigidity, and IPR as a function of boundary parameters. \textbf{e,f} Phase rigidity (top) and IPR (bottom) as a function of $t_{\rm b}$ and $\varepsilon$ under GBC \textbf{e} and OBC \textbf{f}. The appearance of EP corresponds to abrupt changes in both the phase rigidity and IPR, indicating topological phase transitions. \textbf{g} Comparison between theoretically (clean) calculated and experimentally measured (disordered) GBZ. \textbf{h}  Experimentally measured GBZs and corresponding eigenstate distributions under OBC (top row) and GBC (bottom row) for different boundary parameters. The emergence of topological states is clearly visualized in complex and real space.}
  \label{figs:fig3}
\end{figure*}

\begin{figure*}[p]
 \centering
  \includegraphics[width=0.9\textwidth]{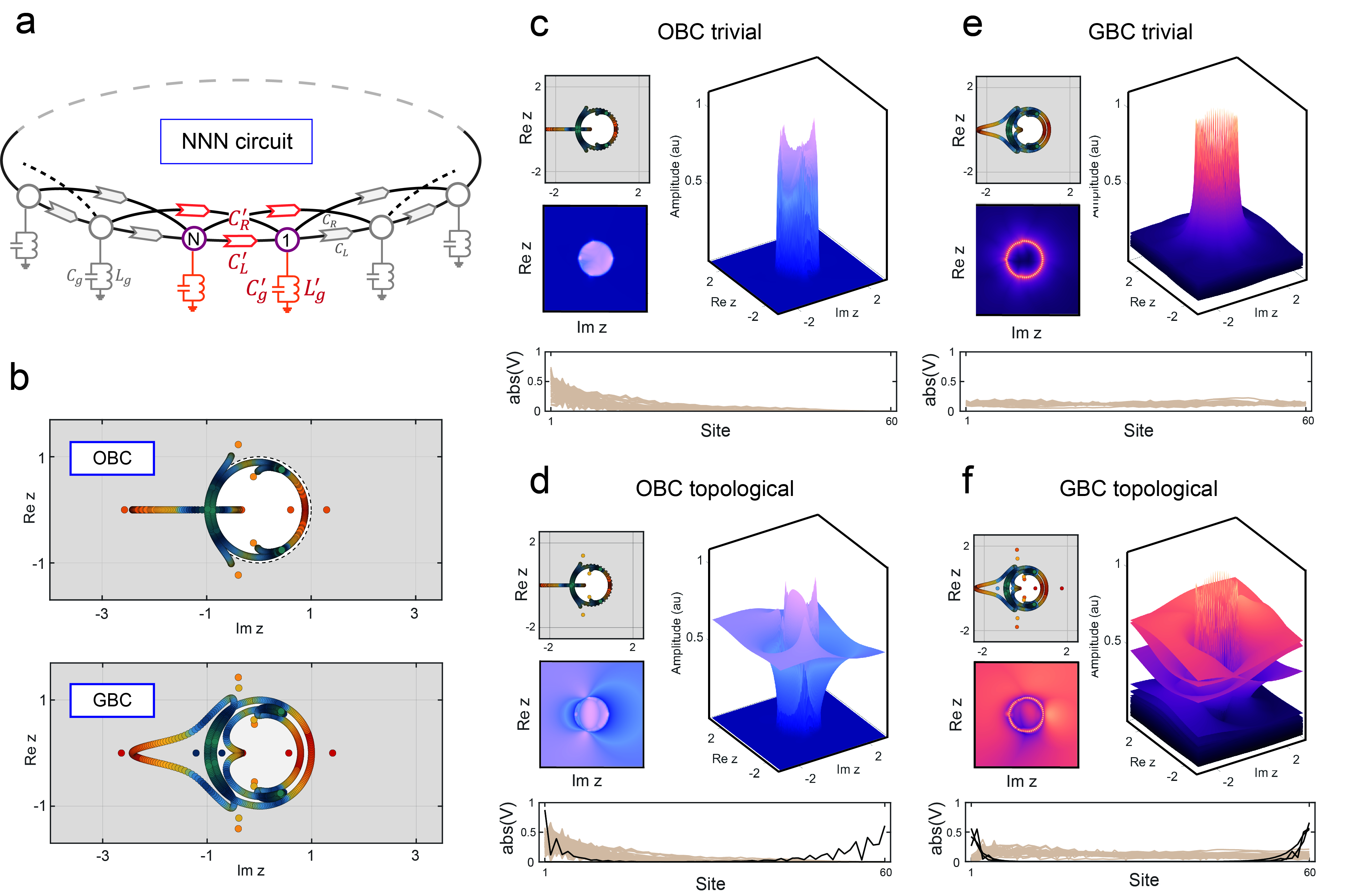}
   \caption{\textbf{Topological phase transition of GBZ in complex plane for NNN case.} \textbf{a} Sketch of the next-nearest-neighbor (NNN) circuit model, showing additional long-range coupling components (INICs).\textbf{b} Theoretical calculation of GBZ under OBC (blue) and GBC for NNN model. In topological phase, multiple topological boundary states appear at complex plane. \textbf{c-f} GBZ structure of NNN model and corresponding eigenstate distributions. For each case, top-left panel shows the theoretical data and bottom-left panel displays a top view of measured GBZ. The right panel presents a 3D visualization of the measured GBZ structure, and the bottom panel shows the eigenstate distribution across lattice sites.}
  \label{figs:fig4}
\end{figure*}

\end{document}


\renewcommand{\thefigure}{S\arabic{figure}}
\setcounter{figure}{0}

\clearpage
\newpage

\begin{widetext}
	\appendix

\section*{Supplementary Material}
\tableofcontents

\clearpage
\newpage

\section{Intrinsic topology of generalized Brillouin zone}

\subsection{Non-Bloch band theory}

This section describes the non-Bloch band theory, focusing on the paradigmatic example of non-Hermitian systems: a nearest-neighbor tight-binding model with asymmetric hopping, i.e., the Hatano-Nelson model. The model Hamiltonian can be expressed as follows: 
\\
\bea
\label{eqs:Hamiltonian}
H = \sum_{i=1}^{N-1} \left(t_{\rm R} c^{\dagger}_{i+1} c_i + t_{\rm L} c^{\dagger}_i c_{i+1} \right) + t'_{\rm R} c^{\dagger}_1 c_N + t'_{\rm L} c^{\dagger}_N c_1 + \varepsilon_1 c^{\dagger}_1 c_1 + \varepsilon_N c^{\dagger}_N c_N.
\eea
\\

Here, $t_{\rm R}$ and $t_{\rm L}$ denote the bulk hopping amplitudes to the right and left, respectively, while $t'_{\rm R}$ and $t'_{\rm L}$ represent the boundary hopping amplitudes to the right and left. The parameters $\varepsilon_{1,N}$ represent the onsite potentials at the boundary nodes. We define the boundary conditions as follows:
\begin{itemize}
    \item Open boundary condition (OBC): \( t'_{\rm R} = t'_{\rm L}  = 0 \).
    \item Periodic boundary condition (PBC): \( t'_{\rm R} = t_{\rm R} \), \( t'_{\rm L} = t_{\rm L} \), \( \varepsilon_1 = \varepsilon_N = 0 \).
    \item Generalized boundary condition (GBC): Any boundary condition that deviates from OBC or PBC.
    \\
\end{itemize}
Non-Bloch band theory is essential in establishing bulk-boundary correspondence in non-Hermitian systems. It extends conventional Bloch's theorem to non-Hermitian systems, allowing the generalization of Brillouin zone to generalized Brillouin zone (GBZ) contours, which characterize the behavior of eigenstates in the complex momentum plane \cite{PhysRevLett.121.086803,PhysRevLett.123.066404,PhysRevLett.125.126402}. The eigenstates under GBC require careful consideration because the boundary terms can induce bound states, affecting the overall localization properties of the system, which is absent in systems under OBC or PBC. We adopt the method presented in Refs. \cite{verma2024topological, PhysRevLett.127.116801} to solve the eigenvalue equation under GBC. We begin by introducing the bulk equations of our system:
\bea
t_{\rm R} \psi_{n-1}-E\psi_n+t_{\rm L} \psi_{n+1}=0,~n\in{2,...,N-1}
\eea
We solve this equation with ansatz wave function \( \psi_n \propto z_j^n \), assuming the eigenstates are a superposition of exponential components characterized by generalized complex momenta \( z_j \). Substituting this ansatz into the bulk equation yields the characteristic equation:
\\
\begin{align}
   \frac{t_{\rm R}}{z} -E + t_{\rm L} z= 0.
\end{align}
\\
The roots of this equation, \( z_1 \) and \( z_2 \), represent the generalized momenta:
\\
\begin{align}
    z_1 = r e^{i\varphi}, \quad z_2 = r e^{-i\varphi}, \quad r = \sqrt{\frac{t_{\rm R}}{t_{\rm L}}}.
\end{align}
\\
Here, \( r \) is related to the ratio between right and left hopping amplitudes, and \( \varphi \in \mathbb{C} \) is a complex number that encapsulates both the oscillatory phase and the damping or amplification effects in the complex plane.
The general solution to the bulk equation can then be expressed as:
\\
\begin{align}
    \psi_n = c_1 z_1^n + c_2 z_2^n.
    \label{eqs:generalsolution}
\end{align}
\\
where \( c_1 \) and \( c_2 \) are coefficients representing the contributions of \( z_1 \) and \( z_2 \) to the wave function. The constraint \( z_1 z_2 = r^2 \) arises directly from the bulk equation, ensuring consistency with the system's dynamics. By considering the bulk ansatz Eq. \ref{eqs:generalsolution} and boundary terms in Eq. \ref{eqs:Hamiltonian}, the boundary equation can be obtained as:
\\
\begin{equation}
H_{\rm B}\begin{pmatrix}
    c_1 \\ c_2
\end{pmatrix}=0,\
\ \ \ \text{with} \ \ \  
H_{\rm B} = \begin{pmatrix} 
A(z_1) & A(z_2) \\
B(z_1) & B(z_2)
\end{pmatrix}
\label{eqs:bdyeq}
\end{equation}
\\
where 
\\
\bea
A(z)=t_{\rm R}-\varepsilon_1 z-t'_{\rm R} z^N, \\ \nonumber
B(z) = t'_{\rm L} z + \varepsilon_N z^N - t_{\rm L} z^{N+1}.
\label{eqs:ABz}
\eea
\\
Under GBC, the boundary matrix $H_{\rm B}$ captures the influence of the boundary conditions by modifying the hopping amplitudes and onsite potentials at the boundaries, directly affecting both the system's eigenstates and eigenvalues. The paired generalized momenta $(z_1,z_2), ~z_1 z_2 =r^2$ are determined by solving the determinant condition \(\det[H_{\rm B}] = 0\) \cite{PhysRevLett.121.086803,PhysRevLett.123.066404}. This condition ensures that the eigenstates satisfy the bulk and boundary equations, thereby establishing the GBZ that accounts for the effects of nontrivial boundary conditions. The determinant condition serves as a constraint linking boundary effects to bulk wave properties, which is crucial for understanding the system's GBZ topology under GBC. 

\subsection{Non-Bloch topological invariant}
The boundary matrix $H_{\rm B}$ is closely connected to the scattering matrix $S$, which relates the incoming and outgoing wave functions through the boundary conditions (Fig. \ref{figs:figs1}). Specifically, the constraints imposed by $H_{\rm B}$ determine the reflection ($R_{ij}$) and transmission ($T_{ij}$) amplitudes, which are key components of $S$. The scattering matrix approach is detailed in the main text. The boundary matrix $H_{\rm B}$ encapsulates the effects of both bulk and boundary conditions. To further analyze the topological properties of the system, $H_{\rm B}$ can be reformulated as an effective boundary Hamiltonian $\tilde{H}_{\rm B}$, which reduces the boundary problem to an effective eigenvalue problem. The effective Hamiltonian formalism helps derive the winding number, a key topological invariant characterizing the GBZ topology. The effective boundary Hamiltonian can be written as:
\\
\begin{equation}
\tilde{H}_{\rm B} = \begin{pmatrix} 0 & h_{\rm B}^+(z_1,z_2) \\ h_{\rm B}^-(z_1,z_2) & 0 \end{pmatrix},
\end{equation}
\\
where $h_{\rm B}^+(z_1,z_2)$ and $h_{\rm B}^-(z_1,z_2)$ are meromorphic functions of the generalized momenta, defined as $h_{\rm B}^+(z_1,z_2) = A(z_2)/A(z_1)$, $h_{\rm B}^-(z_1,z_2) = B(z_1)/B(z_2)$. The Hamiltonian \( \tilde{H}_{\rm B} \) exhibits chiral symmetry, satisfying the anti-commutation relation \( \{\tilde{H}_{\rm B}, \sigma_z\} = 0 \), where \( \sigma_z \) is the Pauli matrix. Such symmetry is crucial in defining the topological invariant of the system, as it provides a foundation for the quantization of the non-Bloch winding number. This quantization is essential for effectively classifying the topological phases of the system \cite{PhysRevX.9.041015}.

To determine the non-Bloch winding number, we employ Cauchy's argument principle, which relates the number of zeros and poles of a meromorphic function $h_{\rm B}^\pm$ to the contour integration of its logarithmic derivative. Specifically, the winding number of the boundary Hamiltonian is given by \cite{verma2024topological}:
\\
\bea
W_{\pm} = \frac{1}{2\pi i} \oint_{\mathcal{L}_1 \times \mathcal{L}_2} \frac{1}{h_{B}^{\pm}(z_1, z_2)} \frac{d h_{B}^{\pm}(z_1, z_2)}{d z_1} d z_1=\frac{1}{2\pi}[\arg h_B^{\pm}]_{\mathcal{L}_1 \times \mathcal{L}_2}=Z-P
\label{Eqs:winding}
\eea
\\
where $\mathcal{L}_1 \times \mathcal{L}_2$ represents the continuous contour of the GBZs satisfying $z_1 z_2=t_R/t_L$, and $[\arg h_B^{\pm}]_{\mathcal{L}_1 \times \mathcal{L}_2}$ is the change of phase of $h_B^{\pm}$. The terms $Z$ and $P$ denote the number of zeros and poles within the GBZ. The distinct topological phases of the GBZ, corresponding to different winding numbers, are shown in Fig. 1e of the main text. Notably, the winding numbers $W_+$ and $W_-$ are generally unequal. We define the non-Bloch winding number as:
\\
\bea
W_{\text{non-Bloch}} \equiv W= \frac{W_+ - W_-}{2}, W'= \frac{W_+ + W_-}{2}
\eea
\\
However, under the condition $\text{det}[H_{\rm B}] = 0$, $W_+$ is always $-W_-$. Therefore, only $W$ can take nontrivial values. The quantized changes in the non-Bloch winding number describe the topological phase transition of GBZ and correspond to the emergence of boundary modes. These modes are topologically protected by the intrinsic topology of GBZ and can be observed as localized eigenstates at the system boundaries.

\subsection{Scattering matrix approach and non-Bloch band theory}

\begin{figure*}[tb!]
  \centering
  \includegraphics[width=0.75\textwidth]{fig_sm1 (2).jpg}
  \caption{\textbf{(a)} Non-Bloch band theory in the Hatano--Nelson (HN) model with boundary hoppings 
  \(\bigl(t'_R,t'_L\bigr)\) and boundary onsite potentials \(\bigl(\varepsilon_1,\varepsilon_N\bigr)\). 
  The eigenstates \(\Psi_{L}(z)\) and \(\Psi_{R}(z)\) on each side of the boundary are superpositions of
  modes \(z=e^{ik}\) with \(\lvert z\rvert\neq 1\). 
  \textbf{(b)} Scattering interpretation, where the boundary of the HN model acts as a ``scatterer'' that couples 
  left- and right-propagating waves. The scattering matrix \(\hat{S}\) encodes the reflection and transmission amplitudes 
  \(\{T_{ij},R_{ij}\}\). Poles of these amplitudes (or equivalently zeros of \(\det [H_{\rm B}]\)) signal the emergence 
  of boundary-localized (bound) states.}
  \label{figs:figs1}
\end{figure*}

\vspace{1em}

\noindent
\textbf{Scattering Matrix Approach.}
Consider the boundary region of Fig.~\ref{figs:figs1}(b) as an effective scatterer between two semi-infinite leads.  
A wave function \(\Psi_{L}(z)\) incident from the left may be transmitted to the right (\(\Psi_{R}(z)\)) or reflected 
back to the left, and similarly for a wave arriving from the right. Grouping the incident and outgoing amplitudes 
into vectors \(\boldsymbol{\Psi}_{\textrm{in}}\) and \(\boldsymbol{\Psi}_{\textrm{out}}\), the scattering process is
\begin{equation}
   \boldsymbol{\Psi}_{\textrm{out}} \;=\; \hat{S}\,\boldsymbol{\Psi}_{\textrm{in}}, 
   \qquad
   \hat{S} \;=\; 
   \begin{pmatrix}
       T_{11} & R_{12}\\[4pt]
       R_{21} & T_{22}
   \end{pmatrix},
   \label{eq:Smatrix-supp}
\end{equation}
where \(T_{ij}\) (\(R_{ij}\)) are transmission (reflection) amplitudes for waves incident from side \(j\).  

\vspace{5pt}
\noindent
\emph{Poles and bound states.\;}  
A boundary-localized (or bound) state appears if there exists a nontrivial solution to 
\(\boldsymbol{\Psi}\neq 0\) even when the incident wave is absent, i.e.\ \(\boldsymbol{\Psi}_{\textrm{in}}=0\).  
Physically, this means the boundary can ``trap'' a mode without any external drive.  
From Eq.~\eqref{eq:Smatrix-supp}, if \(\boldsymbol{\Psi}_{\textrm{in}}=0\) then we must have 
\(\boldsymbol{\Psi}_{\textrm{out}}=\hat{S}\,\boldsymbol{\Psi}_{\textrm{in}}=0\).  
A non-zero \(\boldsymbol{\Psi}\) cannot satisfy this unless \(\hat{S}\) becomes singular.  
This standard criterion from quantum scattering theory thus identifies bound states with poles of the 
reflection/transmission amplitudes.  
Hence, a pole in \(\hat{S}\) (i.e.\ \(\det \hat{S}=0\)) characterizes the formation of boundary-localized bound states.

\vspace{1em}
\noindent
\textbf{Connection Between the Scattering Matrix and Boundary Matrix Approaches.} After defining the incoming/outgoing wavefunction amplitudes 
$\lvert\Psi_L(z)\rangle$, $\lvert\Psi_R(z)\rangle$ on each side of the boundary, 
we can assemble them into a single matrix relation that incorporates both 
the bulk modes $\psi_n \propto z^n$ and the boundary terms 
$\bigl(t'_R, t'_L, \varepsilon_1, \varepsilon_N\bigr)$. 
Concretely, we find:
\\
\begin{equation}
\label{eq:BoundaryScatteringRelation}
\begin{pmatrix}
   \lvert \Psi_R(z_1)\rangle\\
   \lvert \Psi_L(z_2)\rangle
\end{pmatrix}
=
\begin{pmatrix} 
   z_1 & 0 \\
   0 & z_2^N
\end{pmatrix}
\begin{pmatrix} 
   A(z_1)+1 & A(z_2) \\
   B(z_1) & B(z_2)+1
\end{pmatrix}
\begin{pmatrix} 
   z_1^N & 0 \\
   0 & z_2
\end{pmatrix}^{-1}
\begin{pmatrix}
   \lvert \Psi_L(z_1)\rangle\\
   \lvert \Psi_R(z_2)\rangle
\end{pmatrix}.
\end{equation}
\\
Here, $A(z)$ and $B(z)$ encode the boundary modifications to the bulk equations.
This relation shows explicitly how right/left-moving modes match across the boundary,
and thus how bound states can arise (through poles in the ratio $\lvert \Psi_R\rangle/\lvert \Psi_L\rangle$).
In the boundary matrix approach, this same phenomenon corresponds to $\det[H_{\rm B}]=0$.
Hence, the scattering matrix and boundary matrix descriptions are equivalent:
both predict boundary-localized states when the relevant amplitude ratio diverges.

\vspace{1em}
\noindent
\textbf{Boundary Matrix Approach.}
In the non-Bloch band framework, one can also impose boundary continuity directly in a boundary matrix $H_{\rm B}$ to find
$\det[H_{\rm B}]=0$ as the condition for bound states. Writing $\psi_n=c_1 z_1^n + c_2 z_2^n$ and matching amplitudes yields the ratio ${c_2}/{c_1}\!=\!-\frac{A(z_1)}{A(z_2)}\!=\!- \frac{B(z_1)}{B(z_2)}$, which diverges precisely when $\det[H_{\rm B}]=0$. The emergence of bound states in this approach thus mirrors the pole condition in the scattering matrix. The  ratio \({c_2}/{c_1}\) can be viewed as the ``outgoing'' amplitude (e.g.\ on the right) over the 
``incoming'' amplitude (on the left).  Consequently, its poles directly signify bound states, mirroring the scattering 
view that a bound state arises from a pole in reflection or transmission.  
When \({c_2}/{c_1}\) diverges, one obtains a finite solution \(\psi_n\) even though no ``input'' wave is present 
(\(c_1\to 0\) and \(c_2\to\infty\) in a compensating manner).  
Hence, the zeros/poles of \(\,A(z)\) and \(\,B(z)\) coincide with the emergence of boundary-localized states, precisely paralleling the pole condition of \(\hat{S}\) in the scattering approach.

\vspace{1em}
\noindent
\textbf{Equivalence and Topological Significance.}
Overall, the boundary matrix $H_{\rm B}$ and scattering matrix $\hat{S}$ furnish two equivalent views 
of boundary-localized physics in non-Hermitian systems. Poles in $\hat{S}$ align with zeros of $\det[H_{\rm B}]$, 
and thus both signal bound states. This equivalence also plays a central role in determining non-Bloch topological invariants, 
where the counting of zeros and poles within the GBZ contour sets the winding number 
and predicts topological boundary modes.

\subsection{Phase rigidity and inverse partition ratio (IPR)}

In Hermitian systems, eigenvectors are orthonormal, forming a complete and orthogonal basis. In contrast, a non-Hermitian Hamiltonian $H$ (satisfying $H \neq H^\dagger$) generally exhibits complex eigenvalues and possesses two sets of eigenvectors for each eigenvalue $E_n$:
\\
\begin{equation}
H \lvert \psi_n^{\mathrm{R}} \rangle = E_n \lvert \psi_n^{\mathrm{R}} \rangle, 
\quad
\langle \psi_n^{\mathrm{L}} \rvert H = E_n \langle \psi_n^{\mathrm{L}} \rvert.
\end{equation}
\\
These right $\lvert \psi_n^{\mathrm{R}} \rangle$ and left $\langle \psi_n^{\mathrm{L}} \rvert$ eigenvectors constitute a \textit{biorthogonal} set,
\\
\begin{equation}
\langle \psi_m^{\mathrm{R}} \,\big\lvert\, \psi_n^{\mathrm{L}} \rangle = \delta_{mn},
\end{equation}
\\
which reduces to the usual orthonormality when $H$ is Hermitian.
This biorthogonality condition generalizes orthogonality in non-Hermitian systems. Phase rigidity $R$ quantifies the deviation from orthogonality between the left and right eigenvectors corresponding to the same eigenvalue. It is defined as:
\\
\begin{equation} R_n = \frac{\langle \psi^{\rm R}_n | \psi^{\rm L}_n \rangle}{\langle \psi^{\rm R}_n | \psi^{\rm R}_n \rangle}, \quad n=1,2,\ldots,N \end{equation}
\\
In Hermitian systems, the left and right eigenvectors coincide, making $\langle \psi_n^{\mathrm{R}} \vert \psi_n^{\mathrm{L}} \rangle = 
\langle \psi_n^{\mathrm{R}} \vert \psi_n^{\mathrm{R}} \rangle$, so $R_n = 1$.  In non-Hermitian systems, $R_n$ generally lies in the interval $[0,1]$, reflecting the extent of non-Hermiticity. A $R_n$ value less than 1 indicates a degree of non-Hermiticity, with a value of 0 corresponding to an exceptional point (EP), where both eigenvalues and their eigenvectors coalesce, breaking the usual biorthogonality. At these EPs, $R$ tends to zero ($R_i \to 0$), signaling an exceptional transition of the non-equilibrium system \cite{fruchart2021non}. This transition is associated with the emergence or disappearance of topological boundary states in our system.

The inverse participation ratio (IPR) quantifies the localization of eigenstates, particularly the topological boundary states in our system. It is defined as:
\\
\begin{equation} \text{IPR}_n = \frac{\sum_{\mathbf{r}} | \psi_n(\mathbf{r}) |^4}{\left( \sum_{\mathbf{r}} | \psi_n(\mathbf{r}) |^2 \right)^2}, \quad n=1,2,\ldots,N \end{equation}
\\
where $\psi_i(\mathbf{r})$ represents the wave function of the $i$-th eigenstate at position $\mathbf{r}$. The IPR measures how localized an eigenstate is in real space. A higher IPR value indicates stronger localization, while a lower value suggests a more delocalized state. In our system, as the $R_n$ decreases toward zero at the EP, a topological phase transition occurs, giving rise to localized boundary states. This localization is evidenced by the IPR converging to 1, as observed in both theoretical and experimental results (Fig.~\ref{figs:figs2} and main text Fig. 2). Hence, the vanishing of $R_n$ and the simultaneous increase of the IPR affirm that the coalescence of eigenstates at the EP is accompanied by the emergence of topologically protected, highly localized states.

\begin{figure*}[t!]
  \centering
  \includegraphics[width=0.95\textwidth]{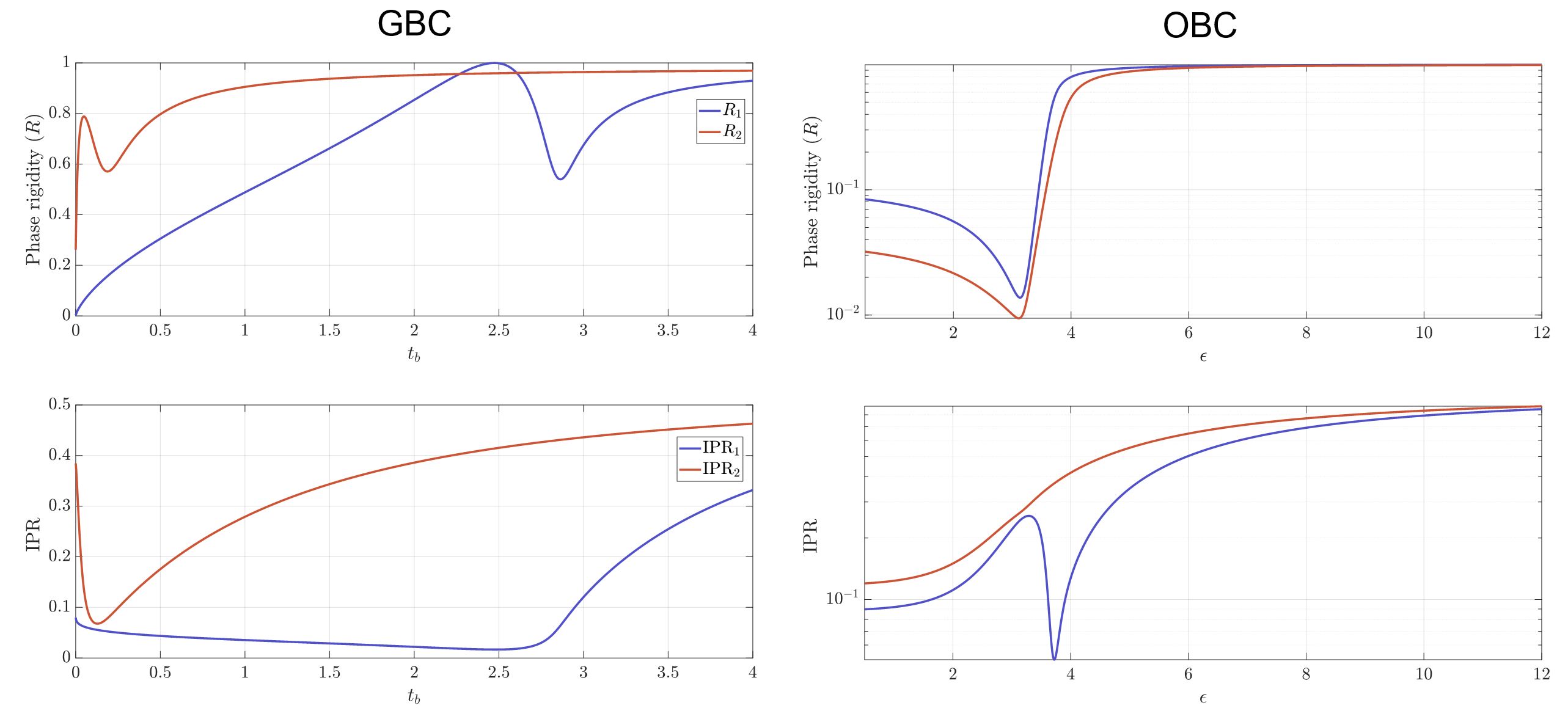}
  \caption{\textbf{Calculation of PR and IPR of topological boundary states under GBC and OBC.} The top panels show the phase rigidity ($R_1$, $R_2$) as a function of boundary parameter $t_{\rm b}$ for GBC and boundary potential parameter $\epsilon$ for OBC. The lower panels plot the corresponding inverse participation ratio (IPR) values. The dips in $R_1,R_2$ and IPR peaks indicate exceptional points (EPs), where eigenstates coalesce, and topological phase transitions occur.}
  \label{figs:figs2}
\end{figure*}

\section{Derivation of circuit Laplacian} 
\subsection{Circuit Laplacian for nearest-neighbor (NN) lattice}
Our model is based on the tight-binding Hamiltonian (see main text Fig. 1a), which is given by:
\\
\bea
H = \sum_{i=1}^{N-1} \left(t_{\rm R} c^{\dagger}_{i+1} c_i + t_{\rm L} c^{\dagger}_i c_{i+1} \right) + t'_{\rm R} c^{\dagger}_1 c_N + t'_{\rm L} c^{\dagger}_N c_1 + \varepsilon_1 c^{\dagger}_1 c_1 + \varepsilon_N c^{\dagger}_N c_N
\eea
\\
where $t_{\rm R} (t_{\rm L})$ is the hopping amplitude for right (left) direction between nearest-neighbor sites, $t'_{\rm R} (t'_{\rm L})$ is the hopping amplitude between boundary sites (main text Fig. 1a), and $\epsilon_1 (\epsilon_N)$ is the onsite potential at the boundary sites, respectively. The circuit Laplacian is defined as the operator $\textbf{I}=J\textbf{V}$ where $\textbf{I}$ is the vector of input currents and $\textbf{V}$ is the vector of response voltages at the nodes.
To derive the circuit Laplacian for our circuit (Fig. \ref{figs:figs3}), we employ Kirchhoff's law on the circuit model, following approach in \cite{lee2018topolectrical}. To illustrate the derivation process, let us first consider a simplified case involving only nodes A and B (dashed box in Fig. \ref{figs:figs3}). In this simplified scenario, the relationship between the input currents $I_A$ and $I_B$, and the corresponding response voltages $V_A$ and $V_B$ at these nodes, can be expressed as:
\\
\bea
\begin{pmatrix} 
I_A \\ 
I_B 
\end{pmatrix} 
=i\omega
\left[
\begin{pmatrix} 
C_1 & -C_1 \\ 
-C_1 & C_1 
\end{pmatrix} 
+
\begin{pmatrix} 
-C_2 & C_2 \\ 
-C_2 & C_2 
\end{pmatrix} 
\right]
\begin{pmatrix} 
V_A \\ 
V_B 
\end{pmatrix}
\eea
\\
where two capacitors used for coupling $C_1,C_2$ are connected with INICs, and we temporarily neglect the grounded LC resonance term for convenience. The asymmetric coupling capacitance, $C_1\pm C_2$, is realized through the use of INICs \cite{PhysRevLett.122.247702}, which effectively create negative impedance. For the entire circuit, the circuit Laplacian can be divided into three parts:
\\
\begin{equation}
J = J_{\text{OBC}} + J_{\text{boundary}} + J_{\text{ground}},
\end{equation}
\\
where \( J_{\text{OBC}} \) is bulk circuit Laplacian assuming open boundary conditions, \( J_{\text{boundary}} \) is the boundary part of circuit Laplacian, and \( J_{\text{ground}} \) is the grounded LC resonance. The bulk circuit Laplacian is given by:
\\
\begin{equation}
\begin{aligned}
J_{\text{OBC}} =& i\omega [(C_1 - C_2 + C_{b1}-\frac{1}{\omega^2L_{b1}})|1\rangle\langle1| + (C_1 + C_2+ C_{bN}-\frac{1}{\omega^2L_{bN}})|N\rangle\langle N| \\
&+ \sum_{x=2}^{N-1} 2C_1 |x\rangle\langle x| - \sum_{x=1}^{N-1} \left((C_1 + C_2)|x\rangle\langle x+1| + (C_1 - C_2)|x\rangle\langle x-1|\right)],
\end{aligned}
\end{equation}
\\
where $C_{b1}$ and $C_{bN}$ are additional capacitors to adjust onsite potential at boundary nodes $1,N$, and $L_{b1}$ and $L_{bN}$ are additional inductors to compensate for any discrepancies in the boundary onsite potentials arising from boundary couplings. The boundary circuit Laplacian is formulated as:
\\
\begin{equation}
J_{\text{boundary}} = i\omega \left[t_{\rm b} \left((C_1 + C_2)|1\rangle\langle1| + (C_1 - C_2)|N\rangle\langle N|\right) - (C_1 + C_2)|N\rangle\langle1| - (C_1 - C_2)|1\rangle\langle N|\right].
\label{eqs:JNbdy}
\end{equation}
\\
where $t_b$ represents a boundary coupling coefficient. The ground term, which considers the resonance effects of capacitances and inductances at each node, is expressed as:
\\
\begin{equation}
J_{\text{ground}} = i\omega \left[\sum_{x=1}^{N} \left(C_g - \frac{1}{\omega^2 L_g}\right)|x\rangle\langle x|\right],
\end{equation}
\\
where \( C_g \) and \( L_g \) denote the ground capacitance and inductance, respectively. By combining the bulk, boundary, and ground contributions, the total circuit Laplacian becomes:
\\
\begin{equation}
\begin{aligned}
J &= J_{\text{OBC}} + J_{\text{boundary}} + J_{\text{ground}} \\
&= i\omega \left[(C_1 - C_2)|1\rangle\langle1| + (C_1 + C_2)|N\rangle\langle N| + \sum_{x=2}^{N-1} 2C_1 |x\rangle\langle x| - \sum_{x=1}^{N-1}\left((C_1 + C_2)|x\rangle\langle x+1| + (C_1 - C_2)|x\rangle\langle x-1|\right)\right] \\
&\quad + t_b \left((C_1 + C_2)|1\rangle\langle1| + (C_1 - C_2)|N\rangle\langle N|\right) - (C_1 + C_2)|N\rangle\langle1| - (C_1 - C_2)|1\rangle\langle N| \\
&\quad + \sum_{x=1}^{N} \left(C_g - \frac{1}{\omega^2 L_g}\right)|x\rangle\langle x|].
\end{aligned}
\end{equation}
\\
This formulation provides the full Laplacian of the circuit, explicitly showing the contributions from bulk, boundary, and ground components. It establishes a direct correspondence between the tight-binding Hamiltonian and the circuit Laplacian, enabling the analysis of topological properties in the electric circuit platform.
\\
\begin{figure*}[t!]
  \centering
  \includegraphics[width=0.95\textwidth]{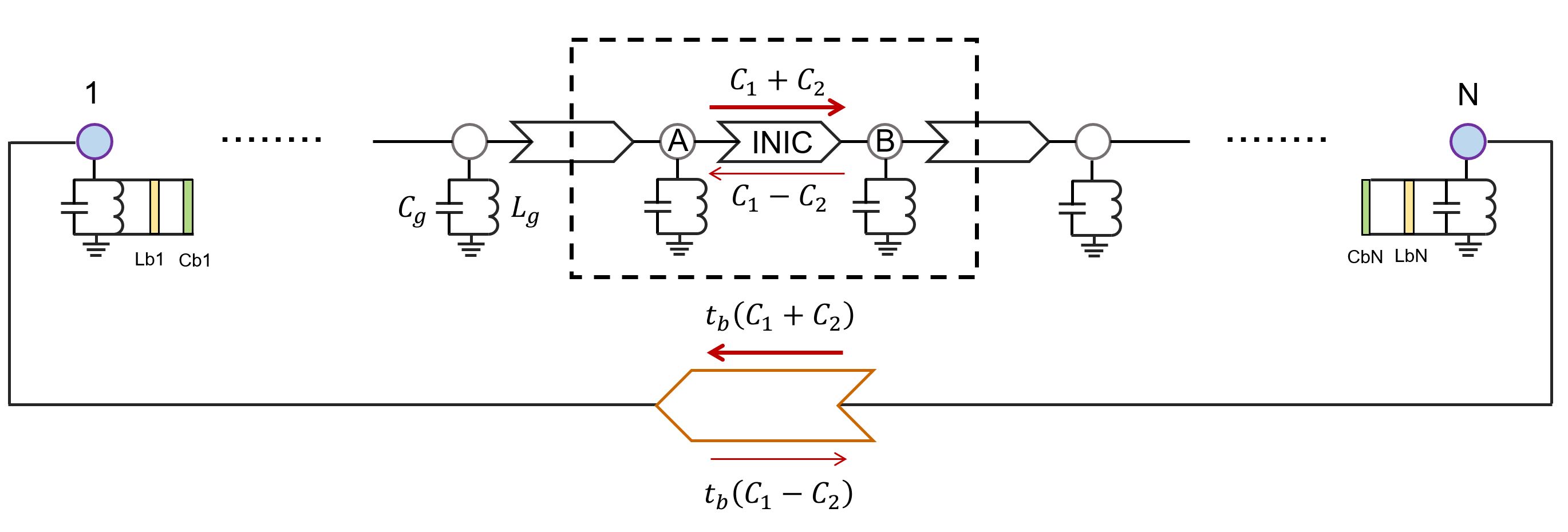}
  \caption{\textbf{Schematic of one-dimensional non-Hermitian circuit model}}
  \label{figs:figs3}
\end{figure*}

\subsection{Circuit Laplacian for next-nearest-neighbor (NNN) lattice}

The one-dimensional Hatano-Nelson model, incorporating next-nearest-neighbor (NNN) interactions, extends the conventional nearest-neighbor framework to capture more complex topological and non-Hermitian phenomena.
Similar to the nearest-neighbor model, the Hamiltonian of the NNN Hatano-Nelson model can be divided into two components: the bulk Hamiltonian, which describes the internal couplings within the lattice, and the boundary Hamiltonian, which describes the interactions at the boundary sites.
\\
\begin{equation}
\hat{H} = \hat{H}_{\text{bulk}} + \hat{H}_{\text{boundary}},
\end{equation}
\\
The bulk Hamiltonian, $\hat{H}_{\text{bulk}}$, describes the couplings between both nearest-neighbor (NN) and next-nearest-neighbor (NNN) sites. It is composed of two sums: one for the nearest-neighbor couplings, and one for the next-nearest-neighbor couplings. The bulk Hamiltonian is given by:
\\
\begin{equation}
\hat{H}_{\text{bulk}} = \sum_{i=1}^{N-1} \left[t_{1\rm R} \hat{c}_i^{\dagger} \hat{c}_{i+1} + t_{1 \rm L} \hat{c}_{i+1}^{\dagger} \hat{c}_i\right] + \sum_{i=1}^{N-2} \left[t_{2 \rm R} \hat{c}_i^{\dagger} \hat{c}_{i+2} + t_{2 \rm L} \hat{c}_{i+2}^{\dagger} \hat{c}_i\right],
\end{equation}
\\
where $t_{1 \rm R},t_{1 \rm L}$ are the right and left hopping amplitudes for nearest-neighbor interactions and $t_{2 \rm R},t_{2 \rm L}$ are the right and left hopping amplitudes for next-nearest-neighbor interactions.
The boundary Hamiltonian, $\hat{H}_{\text{boundary}}$, represents the interactions between the boundary sites of the lattice. The boundary Hamiltonian is expressed as:
\\
\begin{equation}
\hat{H}_{\text{boundary}} = t'_{1 \rm R} \hat{c}_1^{\dagger} \hat{c}_N + t'_{1 \rm L} \hat{c}_N^{\dagger} \hat{c}_1 + t'_{2 \rm R} (\hat{c}_2^{\dagger} \hat{c}_N + \hat{c}_1^{\dagger} \hat{c}_{N-1})+ t'_{2 \rm L} (\hat{c}_N^{\dagger} \hat{c}_2+\hat{c}_{N-1}^{\dagger} \hat{c}_1).
\end{equation}
\\
where $t'_{1 \rm R},t'_{1 \rm L}$ are the right and left hopping amplitudes between the first and last sites of the lattice (node $1$ and node $N$), and $t'_{2 \rm R},t'_{2 \rm L}$ are the right and left hopping amplitudes for NNN interactions involving boundary sites (node $1,2,N-1,N$).

Analogous to the NN setting, we construct the circuit Laplacian for our NNN model (Fig. \ref{figs:figs4}) by applying Kirchhoff's rule to the circuit, integrating both nearest and next-nearest neighbor interactions.
The circuit Laplacian can be divided into three components: the bulk (OBC), the boundary, and the ground circuit Laplacian:
\\
\begin{equation}
J = J_{\text{OBC}} + J_{\text{boundary}} + J_{\text{ground}},
\end{equation}
\\
where $J_{\text{OBC}}$ denotes the bulk Laplacian under OBC, $J_{\text{boundary}}$ relates to the interactions at the boundary (including next-nearest-neighbor interactions), and $J_{\text{ground}}$ includes the ground terms. The comprehensive circuit Laplacian, which includes both nearest neighbor (NN) and next nearest neighbor (NNN) terms, is defined as:
\\
\begin{equation}
\begin{aligned}
J_{\text{OBC}} &= i\omega \Bigg[ 2\left(C_1 - C_2 + C_{b1} - \frac{1}{\omega^2 L_{b1}}\right)|1\rangle\langle 1| + 2\left(C_1 + C_2 + C_{bN} - \frac{1}{\omega^2 L_{bN}}\right)|N\rangle\langle N| \\
&\quad + \left(2C_1 + C_1 - C_2 + C_{b2} - \frac{1}{\omega^2 L_{b2}}\right)|2\rangle\langle 2| +\left(2C_1 + C_1 + C_2 + C_{bN-1} - \frac{1}{\omega^2 L_{bN-1}}\right)|N-1\rangle\langle N-1| \\
&\quad+\sum_{x=3}^{N-2} 4C_1 |x\rangle\langle x| - \left(\sum_{x=1}^{N-1} \left[(C_1 + C_2)|x\rangle\langle x+1| + (C_1 - C_2)|x\rangle\langle x-1|\right]\right) \\
&\quad - \left(\sum_{x=1}^{N-2} \left[(C_1 + C_2)|x\rangle\langle x+2| + (C_1 - C_2)|x\rangle\langle x-2|\right]\right) \Bigg]
\end{aligned}
\end{equation}
\\
where $C_1$ and $C_2$ represent the corresponding capacitances in the NN lattice. The parameters $C_{b1}, C_{b2}, C_{bN-1}, C_{bN}$ represent auxiliary capacitors at the boundary nodes ($1, 2, N-1, N$), whereas $L_{b1}, L_{b2}, L_{bN-1}, L_{bN}$ denote inductors at the same boundary nodes, which avoid unexpected changes in boundary potentials resulting from additional couplings (Eq. B13). The boundary Laplacian is formulated as:
\\
\begin{equation}
\begin{aligned}
J_{\text{boundary}} &= i\omega \Big[t_{\rm {b}} \Big(2(C_1 + C_2)|1\rangle\langle 1| + 2(C_1 - C_2)|N\rangle\langle N| \\
&\quad + (C_1 + C_2)|N-1\rangle\langle N-1| + (C_1 - C_2)|2\rangle\langle 2| \\
&\quad - (C_1 + C_2)|N\rangle\langle 1| - (C_1 - C_2)|1\rangle\langle N| \\
&\quad - (C_1 + C_2)|N\rangle\langle 2| - (C_1 - C_2)|2\rangle\langle N| \Big)\Big]
\end{aligned}
\end{equation}
\\
where $t_{\rm b}$ denotes the boundary coupling coefficient, capturing both NN and NNN boundary interactions. Lastly, the ground term is expressed as:
\\
\begin{equation}
J_{\text{ground}} = i\omega \left[\sum_{x=1}^{N} \left(C_g - \frac{1}{\omega^2 L_g}\right)|x\rangle\langle x|\right]
\end{equation}
\\
where $C_g$ and $L_g$ denote the ground capacitance and inductance, respectively. By combining these components, the total circuit Laplacian becomes:
\\
\begin{equation}
\begin{aligned}
J &= J_{\text{OBC}} + J_{\text{boundary}} + J_{\text{ground}} \\
&= i\omega \left[(C_1 - C_2)|1\rangle\langle1| + (C_1 + C_2)|N\rangle\langle N| + \sum_{x=2}^{N-1} 2C_1 |x\rangle\langle x| - \sum_{x=1}^{N-1}\left((C_1 + C_2)|x\rangle\langle x+1| + (C_1 - C_2)|x\rangle\langle x-1|\right)\right] \\
&\quad + i\omega \left[t_b \left((C_1 + C_2)|1\rangle\langle1| + (C_1 - C_2)|N\rangle\langle N|\right) - (C_1 + C_2)|N\rangle\langle1| - (C_1 - C_2)|1\rangle\langle N|\right] \\
&\quad + \sum_{x=1}^{N} \left(C_g - \frac{1}{\omega^2 L_g}\right)|x\rangle\langle x|.
\end{aligned}
\end{equation}
\\
This formulation explicitly displays the contributions from the bulk, boundary, and ground components, illustrating the direct correspondence between the tight-binding Hamiltonian and the circuit Laplacian.

\begin{figure*}[t!]
  \centering
  \includegraphics[width=0.95\textwidth]{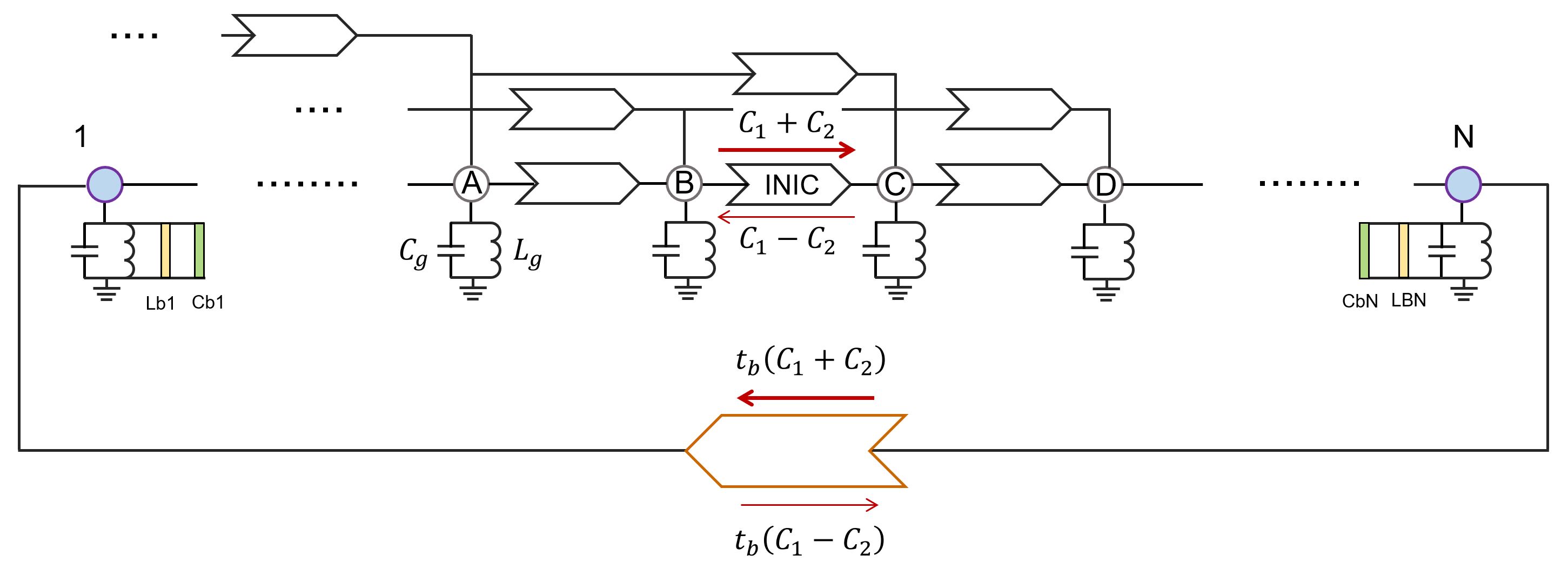}
  \caption{\textbf{Schematic of one-dimensional non-Hermitian circuit model for NNN lattice}}
  \label{figs:figs4}
\end{figure*}

\section{Details of experimental implementation}
For our experimental realization, we fabricated a circuit consisting of 60 nodes on a printed circuit board (PCB). The capacitors (MURATA) with capacitance values of $C_1 = 100 \text{nF}$ and $C_2 = 20 \text{nF}$ were used with operational amplifiers (TEXAS INSTRUMENTS) to realize the asymmetric capacitances $C_1 + C_2 = 120 \text{nF}$ and $C_1 - C_2 = 80 \text{nF}$. The LC resonators, consisting of a capacitor $C_g$ ($70 \text{nF}$, MURATA) and an inductor $L_g$ ($100 \mu\text{H}$, MURATA), were grounded for bulk nodes to ensure the resonance frequency $f = 60$ kHz. At the boundary nodes, we added capacitors $C_{1g}=120\text{nF}$ and $C_{ng}=80\text{nF}$ to guarantee the equivalent onsite potential $2C_1$ (Eq. B4) under OBC. All nodes were grounded by a resistor $R_g$ ($20 \Omega$, WALSIN) to ensure the stability of the operational amplifiers. To manipulate the boundary onsite potentials, additional boundary capacitors were used to achieve effective capacitances $C_{b1}$ and $C_{bN}$ with tunable switches. We connected seven capacitors in parallel with tunable switches at each boundary node, each with values of $2, 4, 8, 16, 32, 64, 128$ nF. Their linear combination allowed us to obtain effective capacitances from $2$ nF to $254$ nF in $2$ nF intervals. To adjust the boundary coupling between node 1 and node N, we connected seven capacitors and a switch in parallel, similar to the boundary nodes, and implemented them with an operational amplifier. This allowed us to realize boundary coupling parameter $t_b$ ranging from $0.2$ to $25.4$ (dimensionless). Due to the additional boundary coupling, the boundary onsite potential changed as shown in Eq. \ref{eqs:JNbdy}. To compensate for this, we installed seven inductors at the boundary nodes to achieve effective inductances $L_{b1}, L_{bN}$, ensuring that the onsite potential at the boundary nodes remained unchanged. To minimize disorder, all circuit components were pre-characterized with an LCR meter, and only those with an error within 2\% were selected.

To reconstruct the circuit Laplacian, we performed admittance band measurements \cite{PhysRevB.99.161114}. By inputting a current $I_j$ at a node $j$, we obtained the impedance by measuring the response voltage $V_i^j$ at a node $i$, which is expressed as:
\\
\bea
G_{i,j}=V_i^j/I_j=J_{i,j}^{-1}
\eea
\\
As the impedance matrix $G$ is the inverse of the admittance matrix $J$, we obtained the eigenvalues and eigenstates of the circuit Laplacian. We applied input current using a function generator (Tektronix) at one node of the circuit and measured the voltage response for all nodes using data acquisition systems (National Instruments). For a system with $N$ nodes, the measurement procedure involved exciting one node and measuring the response voltage from all nodes, repeating this process $N$ times to obtain the matrix $G$. In each of these $N$ measurements, the input current was applied at a different node.

\section{Damped wave Fourier transformation} 
\begin{figure*}[t!]
  \centering
  \includegraphics[width=0.95\textwidth]{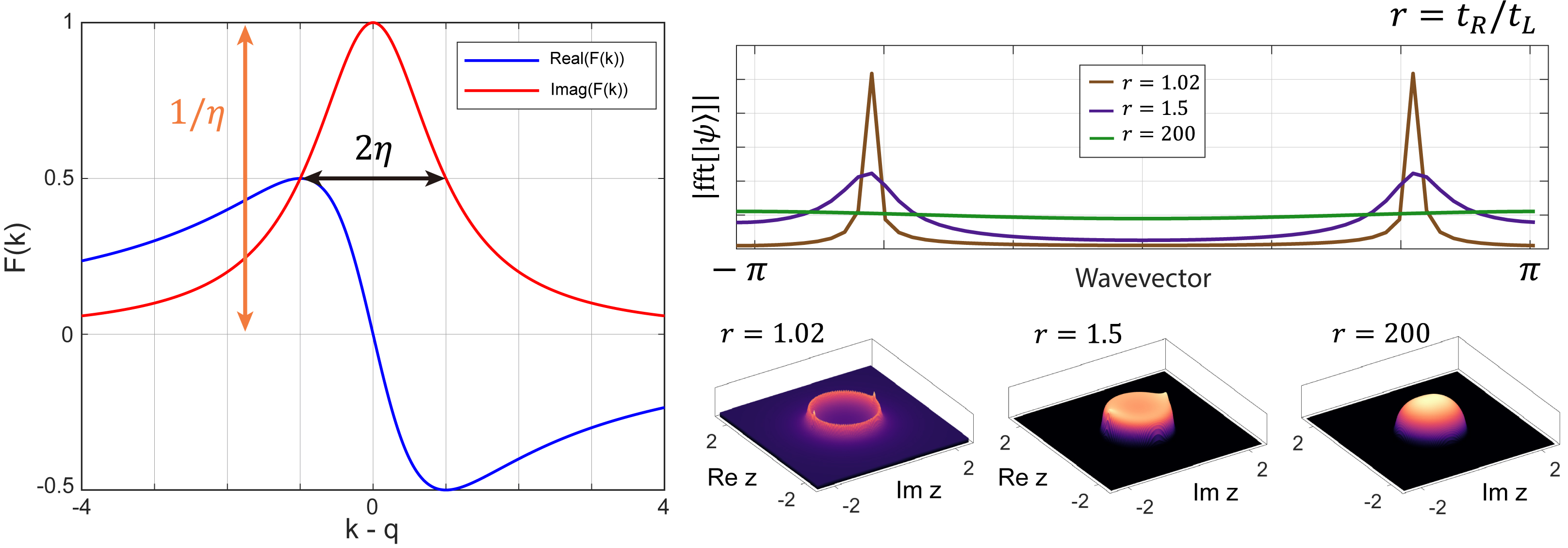}
  \caption{\textbf{Complex Fourier transform and broadening of generalized momenta}}
  \label{figs:figs5}
\end{figure*}
We demonstrate the GBZs in the presence of global disorder using the wave function projection method. This method treats the eigenstates of a finite-size system as linear combinations of incoming, outgoing, damped, or amplified waves. By applying the complex Fourier transform, we extract the generalized complex momenta associated with the eigenstates, where the real part corresponds to the wave vector, and the imaginary part represents the damping or amplification. For example, consider a damped wave described by:
\\
\begin{align}
    f(x) = 
    \begin{cases}
    e^{i(q + i\eta)x}, & \eta > 0, \, x \geq 0, \\
    0, & \eta > 0, \, x < 0,
    \end{cases}
\end{align}
\\
where \( q \) is the real part of the momentum and \( \eta \) is the imaginary part. The Fourier transform of \( f(x) \) is:
\\
\begin{align}
    F(k) = \frac{1}{k - q - i\eta} = \frac{k - q + i\eta}{(k - q)^2 + \eta^2}.
\end{align}
\\
The real part \( q \) determines the peak position of the Fourier transform, while the imaginary part \( \eta \) is characterized by the half-width at half-maximum (HWHM) of the peak. As shown in Fig. \ref{figs:figs5}, a larger imaginary part \( \eta \) corresponds to a broader peak, indicating stronger damping or amplification.

We now apply this Fourier transform approach to the eigenstates of our model described in Eq.~\ref{eqs:Hamiltonian}. As an illustrative example, we consider the bulk eigenstate with eigenvalue \( E_0 = 0 \) under OBC, which satisfies the following bulk equation:
\\
\begin{align}
    t_{\rm R} \psi_{n-1} + t_{\rm L} \psi_{n+1} = 0.
\end{align}
\\
Assuming a plane wave solution \( \psi_n \propto z^n \), substituting into the equation yields:
\\
\begin{align}
    z^2 = -\frac{t_{\rm R}}{t_{\rm L}}.
\end{align}
\\
The solutions are:
\\
\begin{align}
    z = \pm i \sqrt{\frac{t_{\rm R}}{t_{\rm L}}} \equiv \pm ir\equiv = e^{i(\pm\pi/2 - i\log r)},
\end{align}
\\
where \( r = \sqrt{t_{\rm R}/t_{\rm L}} \).
Here, the real part of the momentum \( \text{Re}[k_0] = \pi/2 \) governs the oscillatory behavior of the eigenstate, while the imaginary part \( \text{Im}[k_0] = -\log r \) reflects the damping (or amplification) and the non-Hermitian nature of the system. Consequently, the eigenstate can be expressed as a linear combination of the two solutions \( z_0 \) and \( z_0^{-1} \):
\\
\begin{align}
    \psi_n = c_1 z_0^n + c_2 z_0^{-n},
\end{align}
\\
where \( z_0 = e^{i(\pi/2 - i\log r)} \) and \( z_0^{-1} = e^{-i(\pi/2 - i\log r)} \). The complete eigenstate \( |\psi_{\rm R0}\rangle \) is then written as:
\\
\begin{align}
    |\psi_{\rm R0}\rangle = \sum_n \left( c_1 e^{i(\pi/2 - i\log r)n} + c_2 e^{-i(\pi/2 - i\log r)n} \right) |n\rangle,
\end{align}
\\
where the coefficients satisfy \( c_2 = -c_1 \) due to boundary conditions. The Fourier transform of \( \psi_n \) reveals the real and imaginary parts of the momentum, which correspond to the oscillatory and damping behaviors, respectively.

To determine the generalized momenta associated with the eigenstates of a finite-size system in the presence of global disorder, we numerically diagonalize the Hamiltonian to obtain all eigenvalues and eigenstates. Alternatively, these eigenstates can be experimentally measured. The eigenvalue equation is given by:
\\
\begin{align}
    H |\Psi_n\rangle = E_n |\Psi_n\rangle, \quad n \in \{1, \ldots, N\}.
\end{align}
\\
We construct a two-dimensional grid in the complex plane to represent the generalized momenta:
\\
\begin{align}
    z = \{ x + i y \,|\, x \in [-a, a], \, y \in [-b, b] \}.
\end{align}
\\
At each point \( z \) in this grid, we define the projection operator:
\\
\begin{align}
    P = |\Psi(z)\rangle \langle\Psi(z)|, \quad |\Psi(z)\rangle = (z, z^2, \ldots, z^N)^T.
\end{align}
\\
This operator allows us to evaluate the amplitude of the eigenstate \( |\Psi_n\rangle \) projected onto the \( z \)-dependent basis:
\\
\begin{align}
    S(E_n, \{z_j\}) = |\langle \Psi(z) | \Psi_n \rangle|^2.
\end{align}
\\
The amplitude \( S(E_n, \{z_j\}) \) is peaked around the real part of momenta, while the HWHM of the peak provides an estimate of the imaginary part. For systems with long-range hopping, where \( n_l \)-th neighbors to the left and \( n_r \)-th neighbors to the right are considered, the eigenvalues and eigenstates depend on all \( n_l + n_r \) generalized momenta \( \{ z_j \,|\, j = 1, \ldots, N \} \), which are roots of the bulk polynomial equation:
\\
\begin{align}
    \sum_{l=1}^{n_l} \sum_{r=1}^{n_r} \left( \frac{t_{r\rm R}}{z^r} + t_{l\rm L} z^l \right) - E = 0.
\end{align}
\\
The eigenstates are expressed as:
\\
\begin{align}
    |\Psi_n\rangle = (\psi_1, \ldots, \psi_N)^T, \quad \psi_i = \sum_{j=1}^{n_l+n_r} c_j z_j^i,
\end{align}
\\
where \( c_j \) are coefficients associated with each generalized momentum \( z_j \). For example, in the absence of disorder and under nearest-neighbor hopping, the wave function simplifies to:
\\
\begin{align}
    \psi_i = c_1 z_1^i + c_2 z_2^i, \quad z_1 z_2 = {t_{\rm 1R}}/{t_{\rm 1L}}.
\end{align}
\\
In this case, peak positions of the amplitude \( S(E_n, \{z_j\}) \) determine the real part of the momenta, while the HWHM of the peaks provides an estimate of their imaginary part due to the intrinsic non-Hermiticity. The contributions from disorder (Fig. \ref{figs:figs6}), higher-order hopping, and boundary conditions further influence the peak positions and widths, emphasizing the interplay between bulk and boundary effects in the system.

\begin{figure*}[t!]
  \centering
  \includegraphics[width=0.75\textwidth]{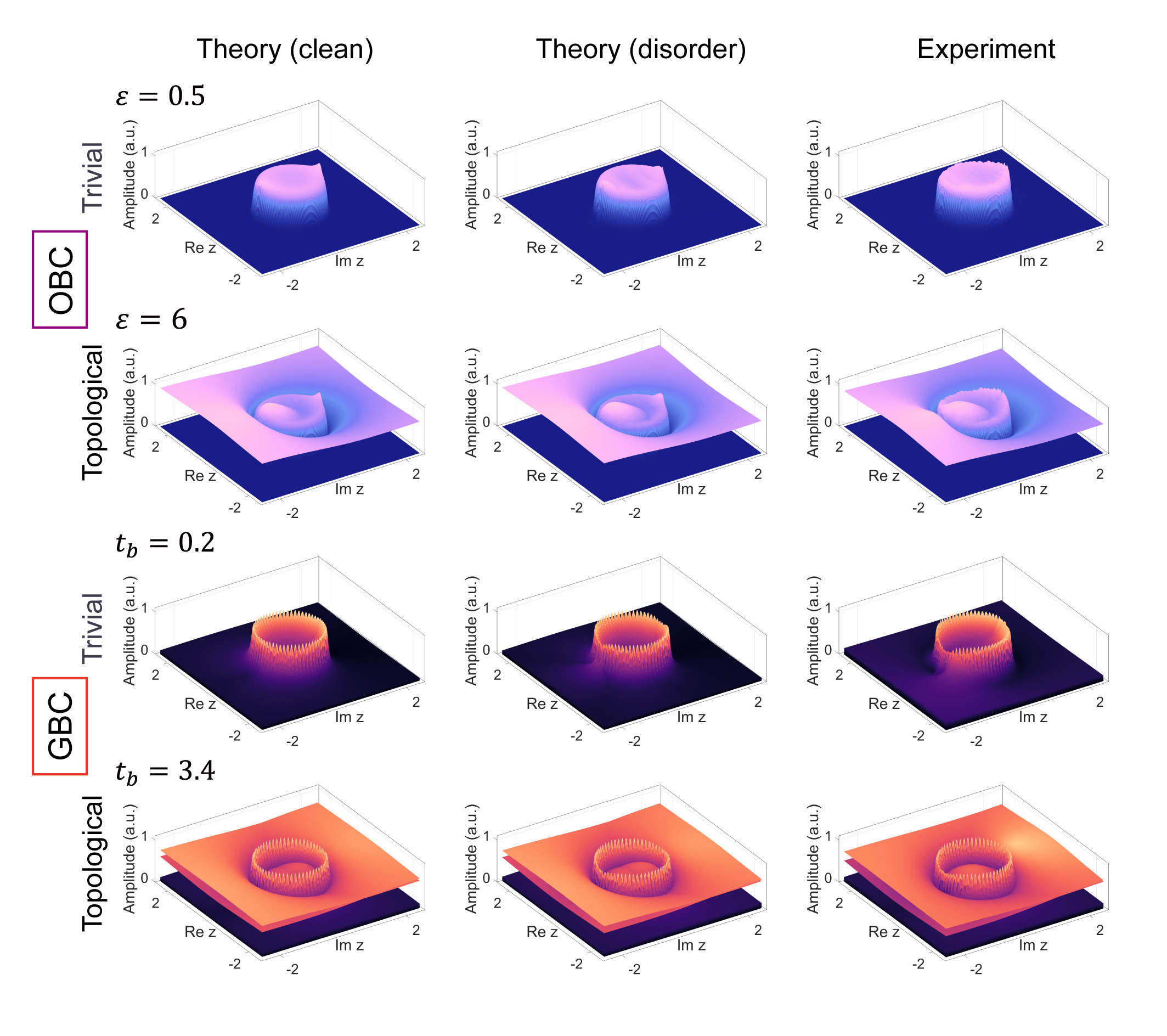}
  \caption{\textbf{Influence of global disorder} Left: Calculated GBZs using wave function projection method without disorder. Middle: Calculated GBZs using wave function projection method with 15\% global disorder. Right: Experimental results. }
  \label{figs:figs6}
\end{figure*}

\end{widetext}

\bibliography{reference}